\documentclass[twocolumn]{aastex631}

\usepackage{amsmath}
\usepackage{amssymb}
\usepackage{bm}
\usepackage{multirow}
\usepackage{etoolbox}
\usepackage[capitalise]{cleveref}

\begin{document}

\title{Two-dimensional Pulsar Distance Inference from Nanohertz Gravitational Waves}

\author{Si-Ren Xiao}
\affiliation{Liaoning Key Laboratory of Cosmology and Astrophysics, College of Sciences, Northeastern University, Shenyang 110819, China}

\author{Ji-Yu Song}
\affiliation{Liaoning Key Laboratory of Cosmology and Astrophysics, College of Sciences, Northeastern University, Shenyang 110819, China}

\author{Yue Shao}
\affiliation{Department of Physics, Liaoning Normal University, Dalian 116029, China}

\author{Ling-Feng Wang}
\affiliation{School of Physics and Optoelectronic Engineering, Hainan University, Haikou 570228, China}

\author{Jing-Fei Zhang}
\affiliation{Liaoning Key Laboratory of Cosmology and Astrophysics, College of Sciences, Northeastern University, Shenyang 110819, China}

\author{Xin Zhang}
\affiliation{Liaoning Key Laboratory of Cosmology and Astrophysics, College of Sciences, Northeastern University, Shenyang 110819, China}
\affiliation{MOE Key Laboratory of Data Analytics and Optimization for Smart Industry, Northeastern University, Shenyang 110819, China}
\affiliation{National Frontiers Science Center for Industrial Intelligence and Systems Optimization, Northeastern University, Shenyang 110819, China}

\correspondingauthor{Xin Zhang}
\email{zhangxin@neu.edu.cn}

\begin{abstract}
Pulsar timing arrays (PTAs) are limited in localizing nanohertz continuous gravitational waves (CGWs) by uncertainties in pulsar distances. We introduce a method to infer pulsar distances in two dimensions, using phase information from the pulsar terms of multiple CGW sources. Our approach can enhance distance precision relative to existing one-dimensional distance-inference methods. Using idealized SKA-motivated PTA simulations with prospective pulsar-distance priors, we combine information from five high-SNR CGW sources with known sky positions and luminosity distances. For example, our method achieves subparsec precision for a simulated pulsar at $0.8\,\mathrm{kpc}$ in 40\% of realizations, compared with none for the existing one-dimensional method. Such improvements in pulsar-distance precision could facilitate CGW host-galaxy identification and multimessenger observational prospects.
\end{abstract}

\section{Introduction}\label{sec:introduction}

Nanohertz gravitational waves (GWs) can be detected with pulsar timing arrays (PTAs), which track pulse times of arrival (TOA) from ensembles of exceptionally stable millisecond pulsars. Passing GWs perturb the Earth-pulsar light-travel time, imprinting characteristic signatures in the timing residuals.

Several PTA collaborations have recently reported evidence for a stochastic gravitational-wave background (SGWB)~\citep{NANOGrav:2023gor,EPTA:2023fyk,Reardon:2023gzh,Xu:2023wog,Miles:2024seg}. The SGWB may arise from various astrophysical and cosmological sources~\citep{Siemens:2006yp,Schwaller:2015tja,Cai:2019bmk,Ellis:2020ena}, with the leading astrophysical contribution expected from the collective GW emission of supermassive black hole binaries (SMBHBs).

In addition to the SGWB, continuous gravitational waves (CGWs) from individual SMBHBs are an equally important PTA target~\citep{Sesana:2008xk,Babak:2011mr,Zhu:2015tua,Wang:2016tfy,Mingarelli:2017fbe}. Identifying their host galaxies, and possibly electromagnetic counterparts, would enable powerful multimessenger studies, such as using nanohertz CGWs to measure the Hubble constant~\citep{Jow:2024bwq,Yan:2020ewq,Wang:2022oou,Jin:2023zhi,Jin:2025dvf} through standard-siren methods~\citep{Schutz:1986gp,Jin:2023sfc,Song:2022siz,Song:2025ddm,LIGOScientific:2025jau,Song:2025bio}. Upcoming observations such as the Legacy Survey of Space and Time (LSST) and the Nancy Grace Roman Space Telescope (Roman) are expected to discover large populations of SMBHB host candidates~\citep{Park:2024gfo,Xin:2025voy,Park:2025eix,Haiman:2023drc}. However, PTAs are expected to localize CGW sources only to ${\sim}\,10^{2}$--$10^{3}\,\mathrm{deg}^{2}$~\citep{Taylor:2015kpa,Goldstein:2018rdr,Petrov:2024hec,Xiao:2024nmi,Truant:2025ybm,Taylor:2026wuz}, limiting host-galaxy identification. This poor localization arises mainly from large uncertainties in pulsar distances, which severely limit the contribution of the pulsar term and lead most studies to treat it as an additional source of self-noise~\citep{Taylor:2014iua,Charisi:2023rdl,Grunthal:2025baq}. Notably, if pulsar distances can be measured with uncertainties below the GW wavelength (${\sim}\,1\,\mathrm{pc}$ at $10\,\mathrm{nHz}$), coherent use of the pulsar terms can significantly improve the sky localization of CGW sources~\citep{Boyle:2010rt,Corbin:2010kt,Lee:2011et,Kato:2025set,Tsai:2026usl,Wen:2026nrc}.

In current PTAs, pulsar-distance uncertainties typically range from tens to hundreds of parsecs, with only a small number of nearby pulsars measured to parsec-level precision through timing or Very Long Baseline Interferometry (VLBI) parallax and kinematic methods~\citep{Deller:2018zxz,Ding:2022luk,Reardon:2024rdv}. Next-generation facilities such as the Square Kilometre Array (SKA) are expected to improve parallax precision to ${\sim}\,15\,\mu\mathrm{as}$ via VLBI imaging~\citep{Smits:2011zh} (compared to the current ${\sim}\,45\,\mu\mathrm{as}$~\citep{Deller:2018zxz}) and to the ${\sim}\,10$--$70\,\mu\mathrm{as}$ level via timing parallax~\citep{Smits:2011zh,Lee:2011et} (compared to the current ${\sim}\,100\,\mu\mathrm{as}$~\citep{NANOGrav:2023hde}). However, parallax-based distance precision degrades rapidly with distance, and most PTA pulsars lie at kiloparsec scales where achieving subparsec distance precision is challenging. These considerations motivate the exploration of additional approaches to constraining pulsar distances.

CGWs provide a complementary means of measuring pulsar distances through the phase difference between the Earth and pulsar terms. For a single CGW source, the phase difference between the Earth term and the pulsar term is measured only modulo $2\pi$, producing a periodic distance degeneracy that prevents precise pulsar-distance inference. \citet{Lee:2011et} showed that combining the CGW phase information with a precise parallax prior can break this periodicity, but this method is effective only for relatively nearby pulsars with tight parallax constraints. More recently, \citet{McGrath:2022cao} and \citet{Yu:2025tnk} showed that multiple CGW sources can constrain pulsar distances. This arises because each source's GW parameters produce distinct periodic structures in pulsar distance posteriors, and mismatched periodicities can suppress the spurious modes in the joint posterior, potentially yielding a dominant peak at the pulsar's true distance.

Developing practical pulsar-distance inference methods with multiple CGW sources remains challenging. In principle, a fully joint analysis across all pulsars retains all inter-parameter correlations without information loss from marginalization, but it becomes computationally intractable for realistic datasets, as the multimodal distance posteriors of many pulsars lead to an exponentially growing parameter space. Existing studies rely on one-dimensional, single-pulsar distance posteriors to infer pulsar distances~\citep{Lee:2011et,Yu:2025tnk}. In particular, when combining information from multiple CGW sources, this one-dimensional approach~\citep{Yu:2025tnk} suppresses the multimodal structure in the joint distance posterior that arises from degeneracies between distance and GW source parameters (such as the chirp mass and initial frequency). This limits the ability of GW observations to constrain pulsar distances.

In this work, we develop a two-dimensional joint-posterior method for pulsar-distance inference using multiple CGW sources, which preserves and exploits the multimodal distance information lost in single-pulsar analyses due to degeneracies between pulsar distances and GW source parameters. Under idealized PTA conditions, assuming the availability of CGW sources with signal-to-noise ratios (SNRs) of 20--40 and known sky positions and luminosity distances, our method can achieve subparsec distance precision for some pulsars and improve pulsar-distance constraints relative to existing one-dimensional approaches. Throughout this work, we use geometrized units with \(G=c=1\).

\section{Simulations and Distance Inference}
\label{sec:simulations}

CGWs from a circular, GW-driven SMBHB induce timing residuals of the form \(s(t,\hat{\Omega})=F^+(\hat{\Omega})\Delta s_+(t)+F^\times(\hat{\Omega})\Delta s_\times(t)\) \citep{Sesana:2010ac,Ellis:2012zv}, where $\hat{\Omega}$ is the GW propagation direction and $F^{+,\times}(\hat{\Omega})$ are the standard PTA antenna-pattern functions. The quantities $\Delta s_{+,\times}(t)$ represent the difference between the Earth and pulsar terms,
\begin{equation}\Delta s_{+,\times}(t)=s_{+,\times}(t_p)-s_{+,\times}(t),\end{equation}
where $t_p=t-L_p(1+\hat{\Omega}\cdot\hat{p})$ is the time at which the pulsar term is evaluated, $L_p$ is the pulsar distance, $\hat{p}$ is the unit vector from the Earth to the pulsar, and $\alpha_p$ denotes the angular separation between the GW source and the pulsar, with $\cos\alpha_p=-\hat{\Omega}\cdot\hat{p}$. Because the pulsar-term signal depends on the phase parameter through $2\Phi_p$, the phase-distance relation yields periodically separated degenerate distance modes. The strain amplitude is given by
\begin{equation}h=\frac{2\,\mathcal{M}^{5/3}\left(\pi f\right)^{2/3}}{d_{\rm L}},\end{equation}
where $\mathcal{M}$ is the redshifted chirp mass, $f$ is the GW frequency, and $d_{\rm L}$ is the luminosity distance. Explicit waveform expressions are adopted from \citet{Becsy:2022zbu}, \citet{NANOGrav:2025gqp}, and \citet{Kato:2025set}.

In our analysis, we assume that the sky locations and luminosity distances of the CGW sources used for pulsar-distance inference are known from external observations and therefore fix these parameters to their injected values during parameter inference. This assumption does not require these quantities to be determined precisely by the PTA itself, since in targeted searches electromagnetic observations can provide the sky position and luminosity distance of an SMBHB candidate, while a CGW detection would associate the GW source with the targeted host galaxy \citep{Jenet:2003ew,NANOGrav:2020lwu,NANOGrav:2025gqp,Tian:2025gjq,Tremblay:2025fgk}. Alternatively, all-sky CGW searches followed by host-galaxy filtering and ranking within the PTA credible sky region can, in favorable cases, reduce the candidates to a single host \citep{Petrov:2024hec,Goldstein:2018rdr,Bardati:2023wbt,Horlaville:2025ebr,Truant:2025ybm}.

Following \citet{Kato:2025set}, we construct a simplified, SKA-era-like 20-yr PTA simulation using 85 pulsars drawn from current PTA datasets. Published distance estimates from the latest PTA releases \citep{NANOGrav:2023pdq,EPTA:2023sfo} and the Australia Telescope National Facility Pulsar Catalogue \citep{Manchester:2004bp} are adopted as fiducial values. We assume a uniform observing cadence of $2\,\mathrm{weeks}$. We adopt an idealized uniform TOA measurement uncertainty of $50\,\mathrm{ns}$ and model the corresponding measurement noise as Gaussian white noise. We also include pulsar-specific intrinsic achromatic red noise, with the red-noise amplitude $A_{\rm RN}$ and spectral index $\gamma_{\rm RN}$ for each pulsar adopted from the corresponding PTA noise analysis \citep{NANOGrav:2023ctt,EPTA:2023sfo,Reardon:2023gzh}. No common red process or stochastic GW background is included in the present simulations.

We generate astrophysically motivated SMBHB populations using \texttt{holodeck} \citep{NANOGrav:2023hfp}, a population-synthesis framework based on semi-analytic models of galaxy mergers and SMBHB evolution, following the approach of \citet{Gardiner:2025mwf}. We adopt the \texttt{Astro\_Strong\_All} model parameter space. The redshifted chirp masses and luminosity distances are derived from the simulated SMBHB populations, while the sky locations, inclinations, polarization angles, and initial phases are randomly sampled. We sample 1000 astrophysical models and generate 100 Poisson realizations for each model. We use the current NANOGrav GWB measurement as an external observational constraint on the SMBHB populations. For each realization, we fit a power-law background to the characteristic-strain spectrum over $1/(16.03\,\mathrm{yr})\leq f\leq10\,\mathrm{nHz}$, approximately corresponding to the first five Fourier bins of the NANOGrav 15-yr data set, with the spectral index fixed to $\gamma=13/3$. We then weight each realization according to the likelihood of its fitted GWB amplitude under the measured value, $A_{\rm GWB}=2.1^{+0.6}_{-0.5}\times10^{-15}$ at a reference frequency of $1\,\mathrm{yr}^{-1}$ \citep{Agarwal:2026dlv}, and draw 20 GWB-conditioned realizations by weighted sampling without replacement. The GWB measurement is used only to condition the SMBHB populations; no stochastic GWB is injected into the simulated PTA data.

\begin{figure}[htbp]
\centering
\includegraphics[width=\columnwidth]{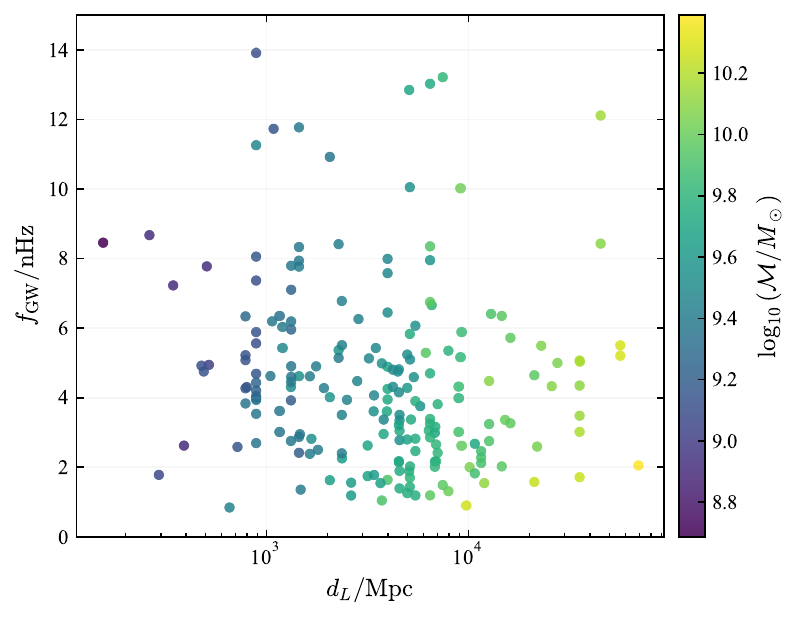}
\caption{Distributions of GW frequency $f_{\rm GW}$, luminosity distance $d_{\rm L}$, and redshifted chirp mass $\mathcal{M}$ for the 200 selected CGW sources with SNRs of 20--40. The color indicates $\log_{10}(\mathcal{M}/M_\odot)$.}
\label{fig_1}
\end{figure}

For each selected realization, we draw the frequency of each CGW source uniformly within its original Fourier bin and compute its optimal network SNR at the drawn frequency using the PTA noise model described above. We then randomly draw $N=5$ or $N=10$ sources without replacement from those with SNRs between 20 and 40. Our results therefore characterize pulsar-distance inference for this SNR-selected source sample. Each CGW source is injected and recovered individually in a separate set of timing residuals. The $N$ data sets within each configuration share the same realization of white and intrinsic red noise. We do not consider the simultaneous superposition or joint recovery of multiple CGW signals. Figure~\ref{fig_1} shows the distributions of GW frequency, luminosity distance, and chirp mass for the 200 selected CGW sources in the $N=10$ configurations.

We perform Bayesian inference on the simulated PTA timing residuals using the standard Gaussian likelihood implemented in \texttt{ENTERPRISE} \citep{enterprise} and \texttt{enterprise\_extensions} \citep{Taylor:2021enterprise_extensions}, with posterior sampling performed using \texttt{PTMCMCSampler} \citep{ptmcmc}. The timing model for each CGW source depends on eight source parameters $(\theta,\phi,d_{\rm L},\iota,\mathcal{M},\psi,\Phi_0,f_0)$, together with the pulsar distance $L_p$ of each pulsar. Here, $(\theta,\phi)$ denote the sky location of the source, $d_{\rm L}$ is the luminosity distance, $\iota$ is the inclination angle, $\mathcal{M}$ is the redshifted chirp mass, $\psi$ is the polarization angle, $\Phi_0$ is the initial GW phase, and $f_0$ is the initial GW frequency. For this proof-of-concept analysis, we fix the intrinsic red-noise parameters to their injected values to reduce the dimensionality of the inference and facilitate convergence. We adopt pulsar-specific timing models from the NANOGrav 15-year \citep{NANOGrav:2023hde}, PPTA DR3 \citep{Zic:2023gta}, and EPTA DR2 \citep{EPTA:2023sfo} data releases for 67, 15, and 3 pulsars, respectively. We marginalize over the linearized timing-model parameters in the likelihood. Consistent with the simulated data, we do not include a common red process or stochastic GWB in the recovery. Although the GW phase offsets at the pulsars, $\{\Phi_p\}$, are not independent parameters, we sample them as free parameters for most pulsars to avoid the highly oscillatory likelihood structure that makes direct sampling in $L_p$ inefficient \citep{Ellis:2013hna}. However, for the five pulsars with subparsec distance-prior uncertainties, we do not treat $\Phi_p$ as an independent free parameter; instead, it is determined by the sampled pulsar distances and GW source parameters through the phase--distance relation, following an approach similar to that of \citet{Wen:2026nrc}. These precise distance priors can provide additional constraints on the GW chirp mass and frequency.

For each pulsar, we impose a Gaussian distance prior with a standard deviation $\sigma_p$ estimated from three approximate SKA-era distance constraints. We estimate the 20-yr timing-parallax precision using the analytic expression of \citet{Lee:2011et}, adopt a fiducial VLBI parallax uncertainty of $15\,\mu{\rm as}$ based on \citet{Smits:2011zh}, and, for binary pulsars with the required orbital-timing measurements available \citep{Egleston:2026mcg}, estimate the potential improvement from an additional 20 years of timing. We adopt the most precise of these estimates as $\sigma_p$. Details are given in Appendix~\ref{app:prior}.

Given the sampled posterior of the GW parameters, we further extract the pulsar-distance information encoded in $\Phi_p$ for the 80 pulsars with free phase parameters using the analytic relation between the pulsar phase parameter and the pulsar distance. Additional details of the phase--distance relation and likelihood structure are provided in Appendices~\ref{app:phip-dp} and \ref{app:likelihood}, respectively. During parameter sampling, this relation is not imposed for these pulsars, thereby avoiding the highly multimodal likelihood associated with phase-linked sampling. We apply this relation only in post-processing, mapping the sampled $\Phi_p$ together with the other relevant GW parameters to the pulsar distance. Because $\Phi_p$ is physically defined only modulo $\pi$, each posterior sample maps to a discrete set of distance solutions corresponding to different integer values of $k$, producing a sequence of $k$-indexed peaks in the resulting distance distribution.

Previous one-dimensional approaches attempted to suppress the spurious $k$-peaks by multiplying the distance posteriors from multiple CGW sources for a single pulsar \citep{Yu:2025tnk}. Different sources impose distinct periodicities on $L_p$, so their posterior product can enhance the true peak. In practice, uncertainties in the GW parameters often wash out the periodicity of these one-dimensional posteriors, limiting the ability of this method to measure the pulsar distance.

We construct a two-dimensional distance posterior for each pulsar pair $(p,q)$ using all $N$ CGW sources,
\begin{equation}\mathrm{PDF}_{pq}(L_p,L_q)\propto\pi_{pq}(L_p,L_q)\prod_{n=1}^{N}\mathcal{P}_{{pq},n}(L_p,L_q).
\label{eq:pair_pdf}
\end{equation}
Here, $\pi_{pq}(L_p,L_q)=\pi_p(L_p)\,\pi_q(L_q)$ is the joint Gaussian prior on the pulsar distances with widths $\sigma_p$ and $\sigma_q$. The quantity $\mathcal{P}_{pq,n}(L_p,L_q)$ represents the contribution of the $n$th CGW source, expressed in the distance space. For pulsars with free phase parameters, this contribution is obtained through the phase--distance mapping. For the five pulsars with subparsec distance priors, we use their sampled distance posteriors, dividing out the distance priors already included in the individual-source posteriors before combining multiple sources. For pulsars with free phase parameters, we neglect the additional pulsar-distance information carried by the frequency differences between the Earth and pulsar terms, as the resulting distance constraints are negligible relative to the pulsar-distance prior. Consequently, we use the pulsar-distance priors when constructing the combined distance distributions. Marginalizing over $L_q$ yields a one-dimensional posterior for pulsar $p$,
\begin{equation}\mathrm{PDF}_p^{(q)}(L_p)=\int dL_q\,\mathrm{PDF}_{pq}(L_p,L_q).\label{eq:marginalized_pdf}\end{equation}
Repeating this procedure for all $q\neq p$ produces a set of posteriors $\{\mathrm{PDF}_p^{(q)}\}$. We adopt the posterior with the smallest half-width of the 68\% equal-tailed credible interval as the final constraint for pulsar $p$.
We define this half-width as the pulsar-distance uncertainty and denote it by $\Delta L_p$.

\section{Pulsar Distance Constraints from CGWs}\label{sec:pulsar_distance_constraints}

\begin{figure}[tbp]
\centering
\includegraphics[width=\columnwidth]{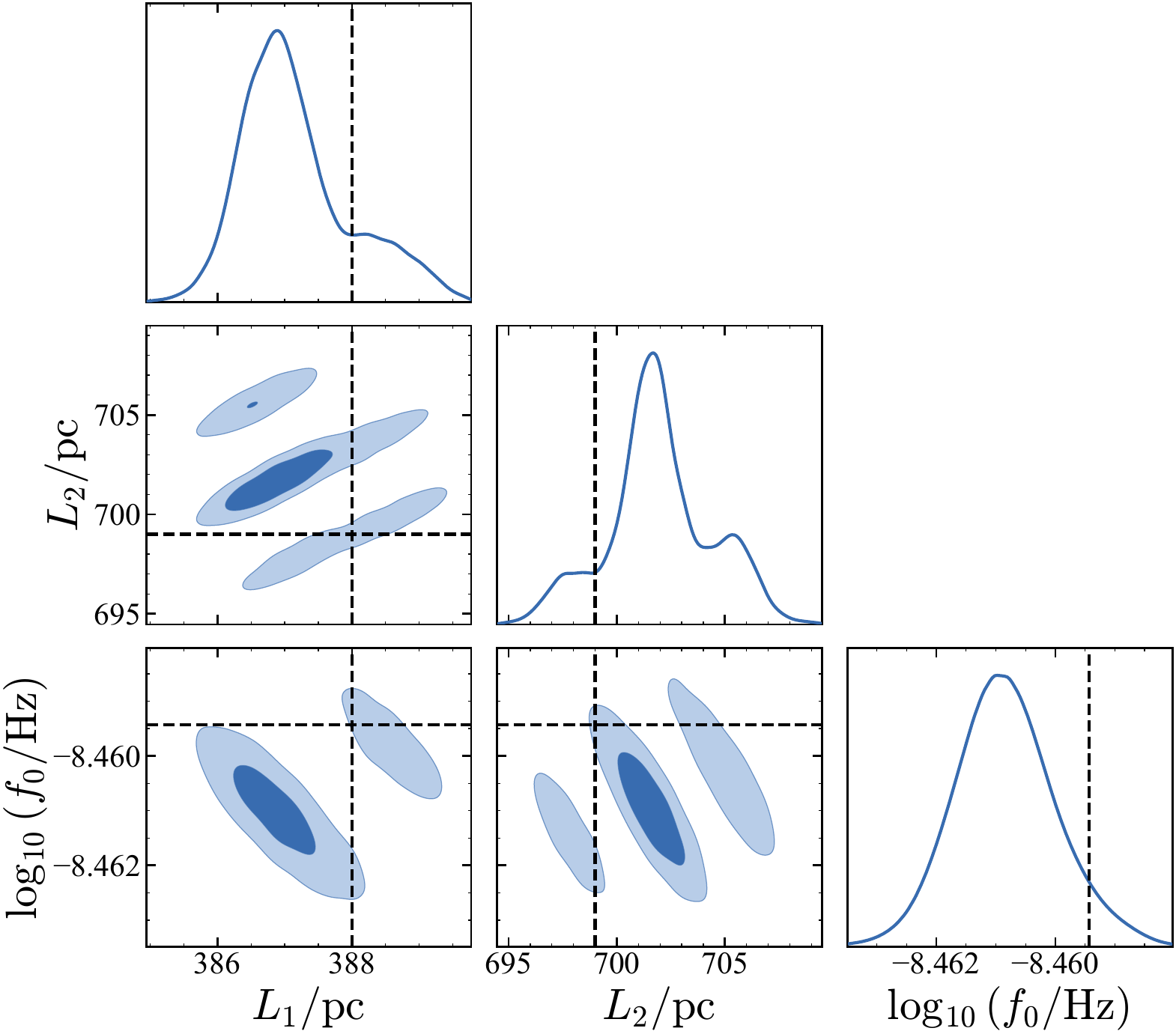}
\caption{Joint posterior distributions of the distances $L_1$ and $L_2$ to pulsars J1744$-$1134 and J1614$-$2230 and the initial GW frequency $f_0$ for a representative GW source. The posterior distributions illustrate how the shared dependence of the pulsar distances on $f_0$ induces correlations between $L_1$ and $L_2$. The contours correspond to the $1\sigma$ and $2\sigma$ credible regions, and black dashed lines indicate the injected values.}
\label{fig_2}
\end{figure}

For an individual GW source, the inferred distances to different pulsars are correlated through their shared dependence on the GW source parameters. Figure~\ref{fig_2} illustrates this mechanism through the joint posterior distribution of the distances $L_1$ and $L_2$ to pulsars J1744$-$1134 and J1614$-$2230, respectively, and the initial GW frequency $f_0$ for a representative source. For each pulsar, the inferred distance is correlated with $f_0$ because the mapping from the pulsar phase to $L_p$ depends on the source frequency. Since both distances depend on the same $f_0$, this induces a correlation between $L_1$ and $L_2$. The dependence of these correlations on the source frequency and sky location is examined in Appendix~\ref{app:correlations}.

\begin{figure}[tbp]
\centering
\includegraphics[width=\columnwidth]{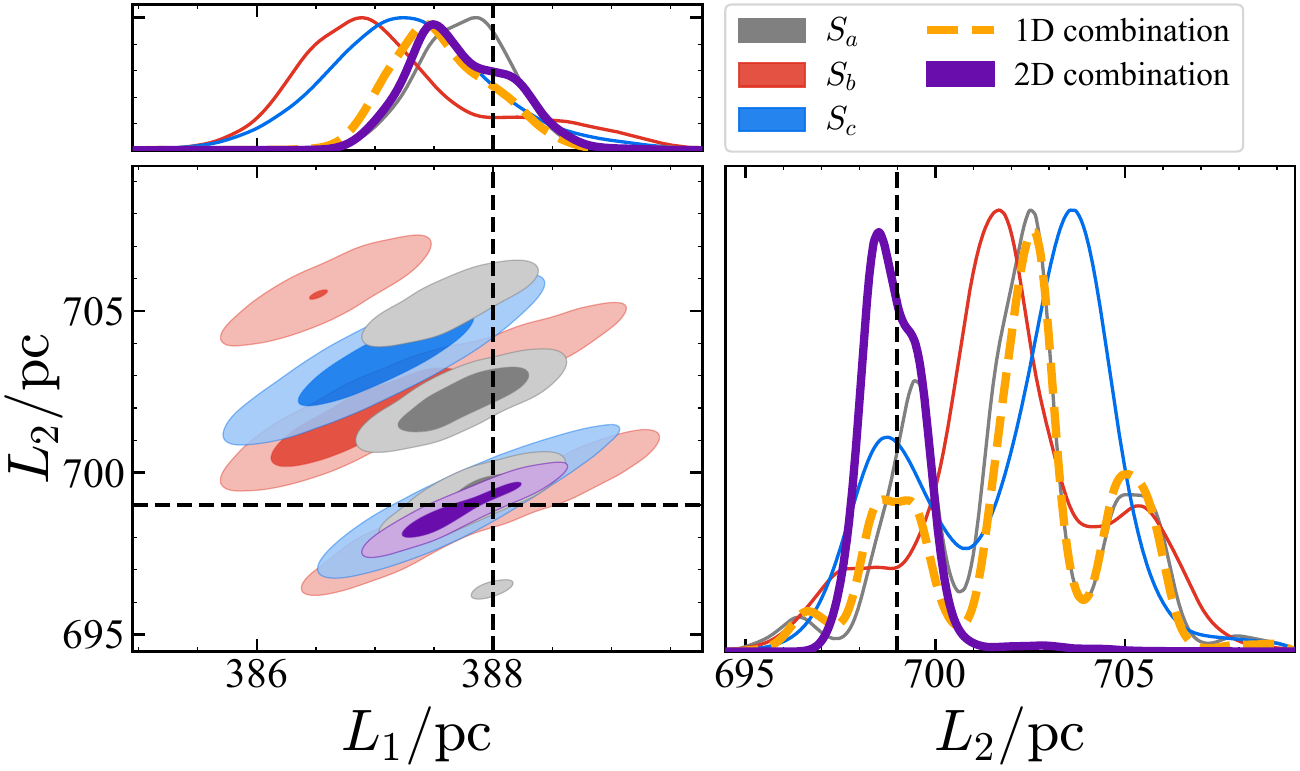}
\caption{Example of pulsar distance inference from the joint two-dimensional posterior obtained by combining three GW sources. Posterior distributions of the distances to pulsars J1744$-$1134 and J1614$-$2230 ($L_1$, $L_2$), including individual posteriors from the three sources and their combined joint posterior (2D combination), are shown. For comparison, we also show the result obtained with the method of \citet{Yu:2025tnk}, which combines the one-dimensional posteriors for each pulsar distance (1D combination). The 1D combination results are shown in orange in the diagonal panels. Black dashed lines indicate the injected true distances. Note that the multimodal structure of J1744$-$1134 is largely suppressed owing to the relatively tight distance prior on this nearby pulsar.}
\label{fig_3}
\end{figure}

To illustrate the improvement obtained by combining pulsar-distance information from separately analyzed GW sources, Fig.~\ref{fig_3} shows the posteriors for the distances to pulsars J1744$-$1134 and J1614$-$2230 from three GW sources in our simulated catalog, as well as their combined constraint. For individual GW sources, the two-dimensional posterior distributions for pulsar distances consist of periodic, approximately parallel degeneracy bands. The spacing of these bands is primarily set by the GW source--pulsar angles $\alpha_p$ and the GW frequency, and therefore differs among sources. When pulsar distance posteriors from multiple GW sources are combined, mismatched bands are suppressed, and the posteriors remain mutually consistent only near the true pulsar distances, yielding a precise pulsar distance measurement.

For comparison, the orange curves in the one-dimensional marginal panels show the results obtained with the existing method, which infers the pulsar distances from one-dimensional marginalized posteriors. This one-dimensional marginalization inevitably projects the distance--distance degeneracies onto a single axis, thereby discarding the correlated two-dimensional structure of the pulsar-distance posterior. Consequently, one-dimensional combinations of multiple GW sources are less effective at isolating the true peak in the pulsar-distance posterior than two-dimensional combinations. In this case, the distance to J1614$-$2230 has a half-width of $1.40\,\mathrm{pc}$ for the 68\% credible interval from the two-dimensional posterior, smaller than the $2.91\,\mathrm{pc}$ obtained with the one-dimensional analysis. Although these results are based on an illustrative example, the statistical performance of our method is evaluated below.

\begin{figure*}[htbp]
\centering
\includegraphics[width=\textwidth]{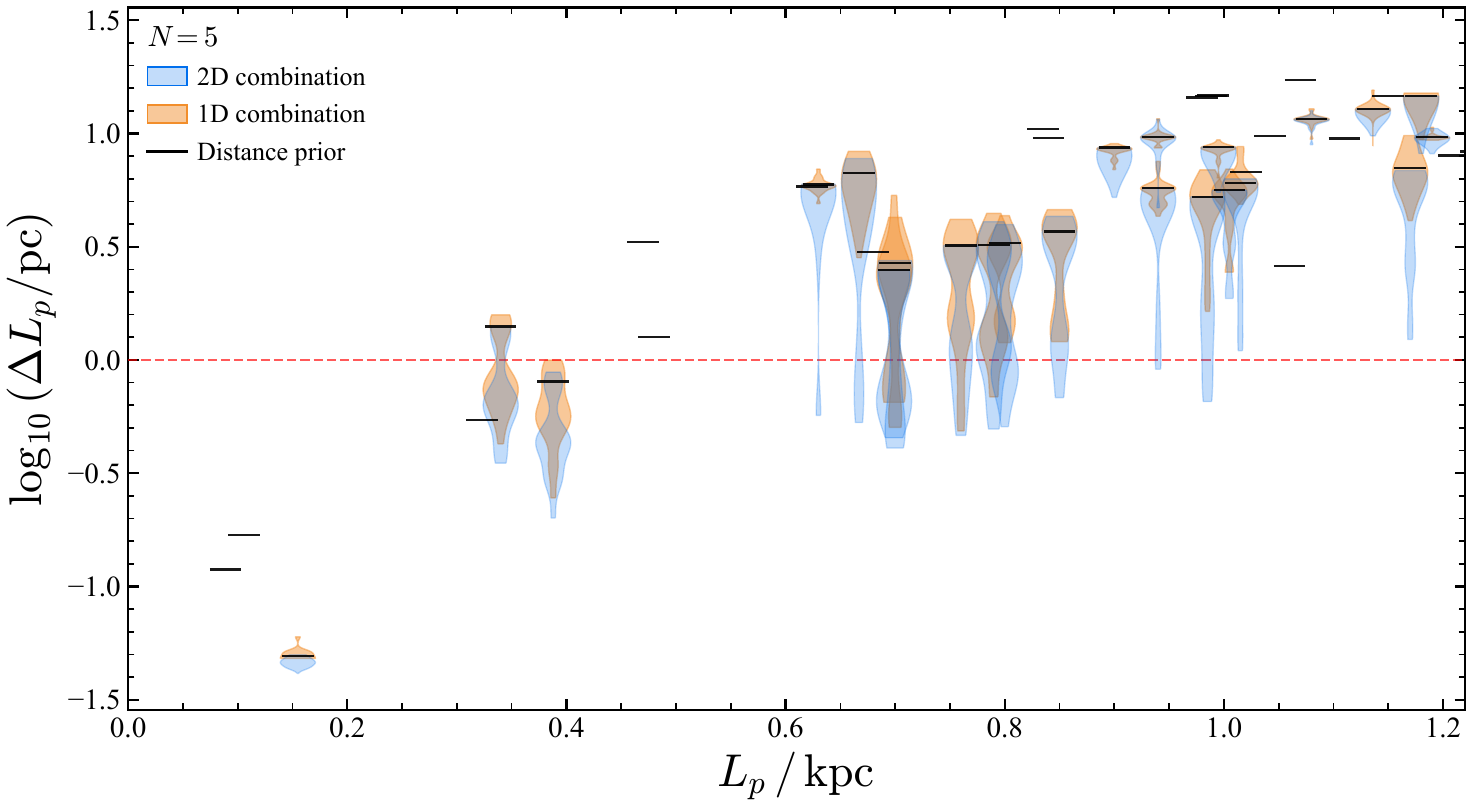}
\caption{Pulsar-distance uncertainties $\Delta L_p$ versus pulsar distance $L_p$ for $N=5$ CGW sources. Violins show the distributions across 20 realizations for the 2D (blue) and 1D (orange) combinations. Black horizontal lines indicate the distance-prior uncertainties, and the red dashed line marks $\Delta L_p=1\,\mathrm{pc}$. For clarity, only pulsars within $1.2\,\mathrm{kpc}$ are shown.}
\label{fig:dp_err}
\end{figure*}

Figure~\ref{fig:dp_err} presents the distributions of the pulsar-distance uncertainty $\Delta L_p$ across 20 realizations of our idealized, SKA-motivated 20-yr PTA simulation, each combining information from $N=5$ CGW sources. The 2D combination typically yields tighter distance constraints than the 1D combination. For J1804$-$2717 ($0.80\,\mathrm{kpc}$) in the simulated array, for example, the median $\Delta L_p$ across the 20 realizations decreases from $2.78\,\mathrm{pc}$ with the 1D combination to $2.02\,\mathrm{pc}$ with the 2D combination. The fraction of realizations achieving subparsec precision increases from 0\% to 40\% (8/20). For some pulsars, strong intrinsic red noise weakens the constraints on the pulsar-term phase, limiting the distance information obtained from CGW signals. Only the distance priors are shown for pulsars without reliable CGW distance constraints. Results for $N=10$ CGW sources are presented in Appendix~\ref{app:ten_sources}.

\begin{figure}[tbp]
\centering
\includegraphics[width=\columnwidth]{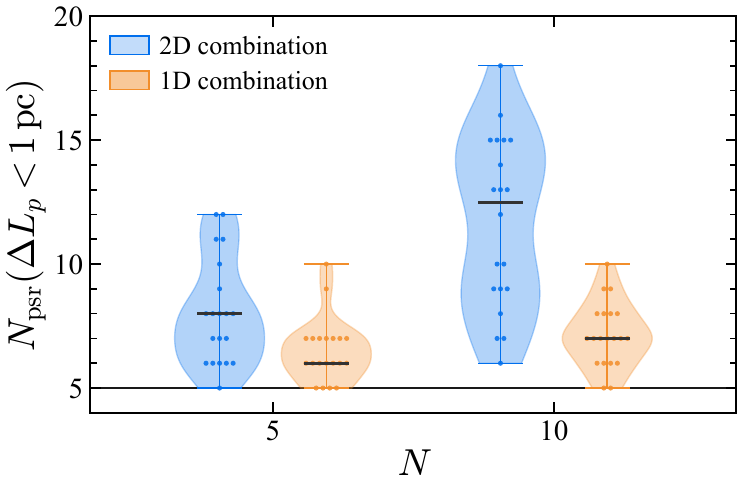}
\caption{Violin plots of $N_{\mathrm{psr}}$ (number of pulsars with $\Delta L_p < 1\,\mathrm{pc}$ in the 85-pulsar array) versus CGW source count $N$, from 20 realizations, for the 2D (blue) and 1D (orange) combinations. Points show individual realizations, and black horizontal lines mark the medians. The dashed line indicates the $N_{\mathrm{psr}}$ obtained with distance priors alone.}
\label{fig_5}
\end{figure}

For the full array, we count how many of the 85 pulsars achieve $\Delta L_p < 1\,\mathrm{pc}$ per realization, as shown in Fig.~\ref{fig_5}. With distance priors alone, only 5 pulsars meet this criterion. Including $N = 5$ and $N = 10$ CGW sources, the median counts with the 2D combination rise to 8 and 12.5, respectively, compared with 6 and 7 with the 1D combination, indicating that combining CGW signals with distance priors yields precise distances for more pulsars than distance priors alone.

\begin{figure}[tbp]
\centering
\includegraphics[width=\columnwidth]{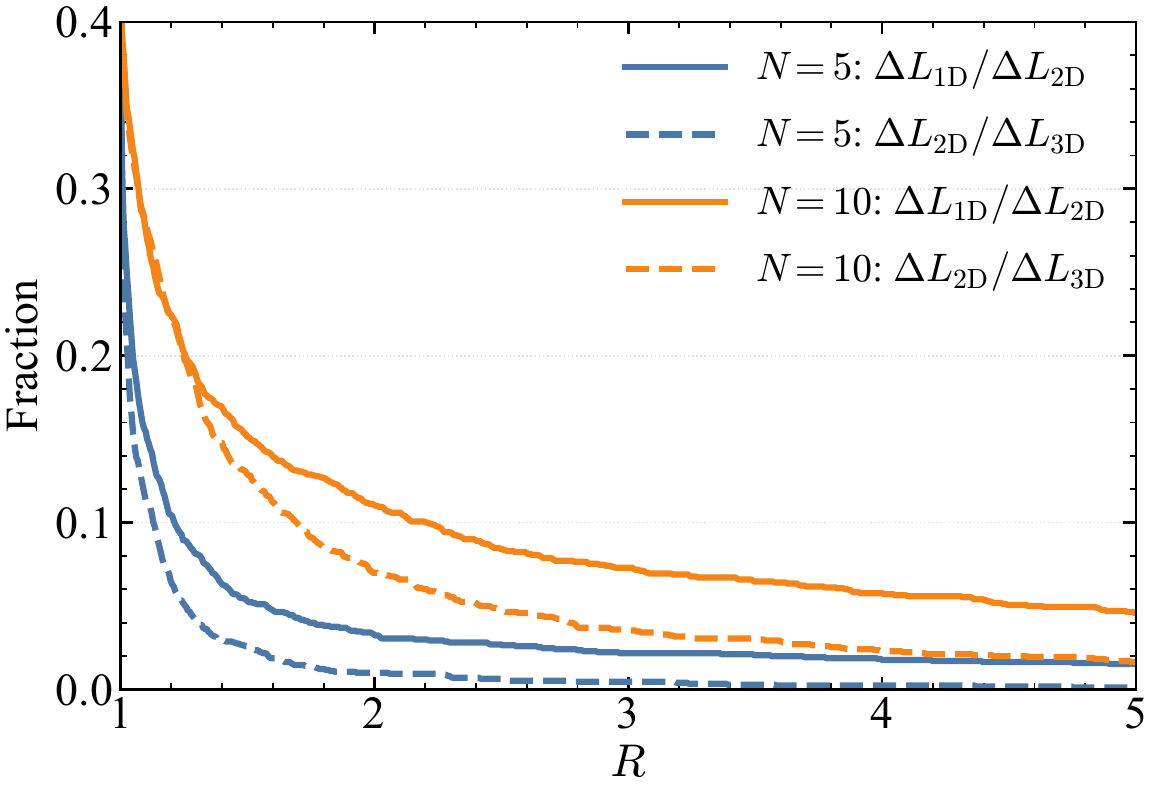}
\caption{Complementary cumulative distributions of pulsar-distance precision improvement factors from 20 independent realizations, each containing 85 pulsars. Blue and orange curves correspond to $N=5$ and $N=10$ independent CGW sources, respectively. Solid curves show $R_{12}=\Delta L_{\rm 1D}/\Delta L_{\rm 2D}$, while dashed curves show $R_{23}=\Delta L_{\rm 2D}/\Delta L_{\rm 3D}$. The vertical axis gives the fraction of samples with improvement factors greater than or equal to $R$.}
\label{fig_6}
\end{figure}

We further extend the 2D combination method to three dimensions and compare the resulting pulsar-distance uncertainties with those obtained from the 1D and 2D combinations. Figure~\ref{fig_6} shows the complementary cumulative distributions of the improvement factors $R_{12}=\Delta L_{\rm 1D}/\Delta L_{\rm 2D}$ and $R_{23}=\Delta L_{\rm 2D}/\Delta L_{\rm 3D}$, based on 20 independent realizations, each containing 85 pulsars. For $N=5$ independent CGW sources, 3.2\% and 1.0\% of the samples achieve improvement factors of at least two for $R_{12}$ and $R_{23}$, respectively. For $N=10$, the corresponding fractions increase to 11.1\% and 7.0\%. These results show that the 3D combination can further improve the distance constraints for some pulsars, while the fraction of samples achieving at least a twofold improvement is lower for $R_{23}$ than for $R_{12}$.

For a PTA comprising 85 pulsars, inferring the distance to a given pulsar involves 84 possible pulsar pairs in the 2D combination and 3486 pulsar triplets in the 3D combination. The 3D combination also requires more computation per combination due to the higher-dimensional joint posterior. Based on the average computation time per combination measured across four realizations with $N=5$, the estimated times required to evaluate all combinations for a single target pulsar are approximately 1.2\,s and 6.3\,min, respectively, corresponding to a factor of about 320 in computational time. These estimates are based on calculations performed on a server equipped with Intel Xeon Silver 4216 CPUs (2.10\,GHz, 64 logical cores). Thus, although combining additional pulsars can provide tighter distance constraints, extending the method to higher dimensions also increases the computational cost.

\section{Conclusions}\label{sec:conclusions}

Previous studies have shown that nanohertz CGWs can provide information on pulsar distances~\citep{Corbin:2010kt,Lee:2011et,McGrath:2022cao,Yu:2025tnk}. Building on these studies, we introduce a method that combines two-dimensional pulsar-distance posteriors inferred from multiple GW signals. By retaining the inter-pulsar distance correlations induced by shared GW parameter uncertainties that are lost in one-dimensional marginalization and avoiding the prohibitive computational cost of a joint inference over all pulsar-distance parameters in the PTA, our approach facilitates pulsar-distance inference based on pulsar-term signals.

Applying our method to an idealized SKA-motivated PTA simulation with white noise and pulsar-specific intrinsic red noise, but without an SGWB or overlapping CGW sources, we find that combining information from multiple CGW sources with SNRs of 20--40 and externally determined sky positions and luminosity distances can achieve subparsec precision in pulsar-distance measurements. The intrinsic red-noise parameters are fixed during parameter recovery. Quantifying the impact of additional GW signals and noise processes on pulsar-distance inference is a natural and necessary extension of this work. In principle, more precise pulsar-distance measurements improve the detectability of additional CGW sources, which in turn enables further distance refinement. A more realistic analysis of such cyclic refinement within a full multiple-source search is left for future work.

Improved pulsar-distance constraints can support coherent use of pulsar-term signals and thereby aid the sky localization of other CGW sources whose positions are less well constrained~\citep{Boyle:2010rt,Lee:2011et,Kato:2025set}. Such improvements could facilitate SMBHB host-galaxy identification and multimessenger follow-up. The actual localization gains, however, depend on the timing sensitivity of individual pulsars, the source--array geometry, and the joint constraints on pulsar distances and phases. Quantifying these gains requires a coherent localization analysis that consistently incorporates the phase--distance relation, which is beyond the scope of the present work.

\begingroup
\let\internallinenumbers\relax
\patchcmd{\acknowledgments}{\vskip 5.8mm plus 1mm minus 1mm}{}{}
  {\PackageError{main}{Unable to adjust acknowledgment spacing}{Check the AASTeX class definition.}}
\patchcmd{\acknowledgments}{\noindent\ignorespaces}{\section*{Acknowledgments}\indent\ignorespaces}{}
  {\PackageError{main}{Unable to indent the acknowledgment paragraph}{Check the AASTeX class definition.}}
\begin{acknowledgments}
We thank Ji-Guo Zhang, Tian-Nuo Li, and Yi-Min Zhang for helpful discussions. This work was supported by the National SKA Program of China (Grants Nos. 2022SKA0110200 and 2022SKA0110203), the National Natural Science Foundation of China (Grants Nos. 12533001, 12575049, 12473001, and 12305058), the China Manned Space Program (Grant No. CMS-CSST-2025-A02), the 111 Project (Grant No. B16009), and the Natural Science Foundation of Hainan Province (Grant No. 424QN215).
\end{acknowledgments}
\endgroup

\bibliography{cgw_dpmeasure}

\clearpage

\appendix
\renewcommand*{\theHequation}{appendix.\Alph{section}.\arabic{equation}}

\section{Likelihood structure in pulsar phase and distance}
\label{app:likelihood}

To illustrate the effect of treating the pulsar-term phase as a free parameter, we perform one-dimensional likelihood scans for PSR J0636$+$5128 using a simulated CGW signal. We adopt a zero-noise injection while retaining the white- and intrinsic-red-noise covariance model used in our analysis. For each scan, all parameters other than the scanned parameter are fixed at their injected values.

\begin{figure}[htbp]
    \centering
    \includegraphics[width=\columnwidth]{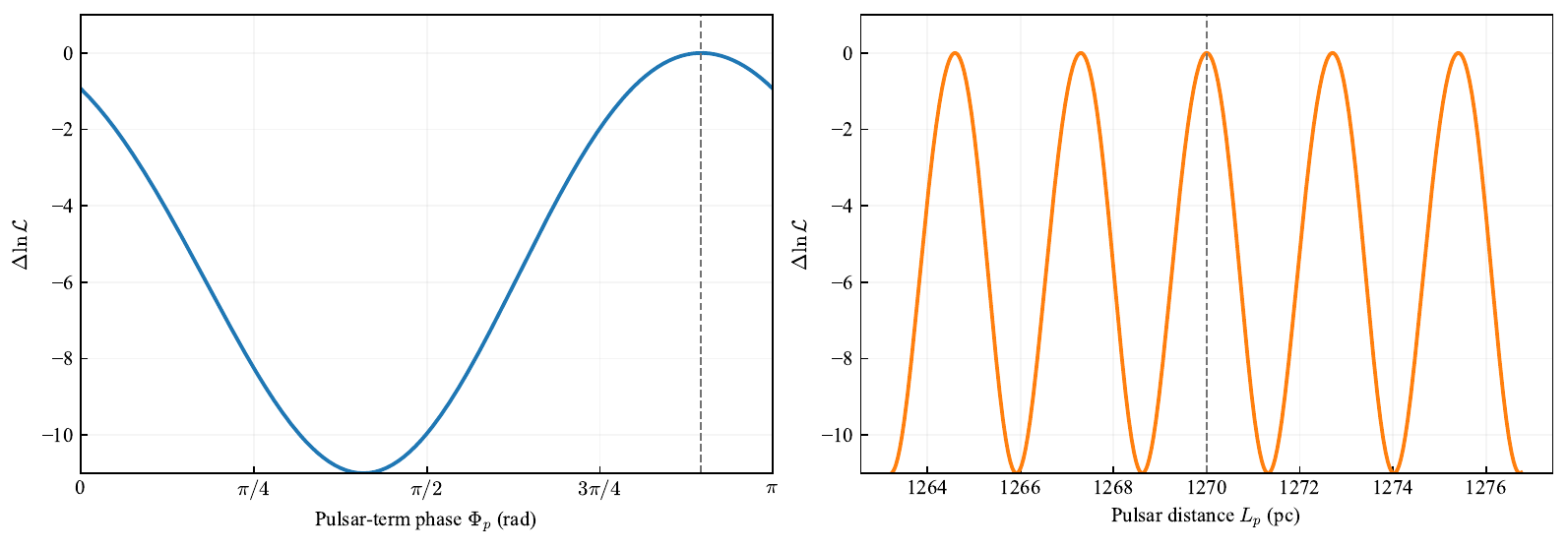}
    \caption{One-dimensional likelihood scans for PSR J0636$+$5128, varying the free pulsar-term phase $\Phi_p$ (left) or the pulsar distance $L_p$ (right). All other parameters are fixed at their injected values. Dashed vertical lines indicate the pulsar-term phase derived from the injected parameters (left) and the injected pulsar distance (right).}
    \label{fig:phase_distance_likelihood}
\end{figure}

Figure~\ref{fig:phase_distance_likelihood} compares the resulting likelihood curves. In the left panel, we vary the pulsar-term phase $\Phi_p$ as a free parameter over one period, without imposing the analytic phase--distance relation. In the right panel, we vary the pulsar distance $L_p$ directly, with the corresponding pulsar-term phase determined by the physical signal model. The distance scan exhibits multiple separated likelihood peaks arising from the periodic dependence of the pulsar-term phase on distance. Treating $\Phi_p$ as a free parameter facilitates MCMC sampling by avoiding direct exploration of these separated distance modes.

\section{Prospective pulsar-distance priors}
\label{app:prior}

To construct approximate distance priors for the prospective 20-yr PTA, we estimate the future precision from timing parallax, VLBI astrometry, and, where applicable, kinematic-distance measurements of binary pulsars.

For timing parallax, we use the analytic estimate of \citet{Lee:2011et},
\begin{equation}
\sigma_{p}^{\rm TP}\simeq
\frac{2.34}{\cos^2\beta_p}
\left(\frac{N_{\rm obs}}{100}\right)^{-1/2}
\left(\frac{L_p}{1\,{\rm kpc}}\right)^2
\left(\frac{\sigma_n}{10\,{\rm ns}}\right)
{\rm pc},
\label{eq:timing_parallax}
\end{equation}
where $\beta_p$ is the ecliptic latitude, $N_{\rm obs}$ is the number of TOAs, and $\sigma_n$ is the white-noise RMS assumed in the original derivation. For a 20-yr observation with a two-week cadence, $N_{\rm obs}\simeq520$.

As an approximate correction for the intrinsic time-correlated red noise included in our simulations, we replace $\sigma_n$ by a single-frequency effective noise level defined by $\sigma_{{\rm eff},p}^{2}=\sigma_{\rm WN}^{2}+S_{{\rm RN},p}(f_{\pi})\,f_{\rm max}$, where $f_{\pi}=2\,{\rm yr}^{-1}$, $f_{\rm max}=1/(2\Delta t)$, $\sigma_{\rm WN}=50\,{\rm ns}$, and $S_{{\rm RN},p}(f)$ is the intrinsic red-noise power spectral density. This correction is used only to estimate the timing-parallax precision.

For VLBI astrometry, we adopt a fiducial future parallax uncertainty of $\sigma_\varpi=15\,\mu{\rm as}$ based on \citet{Smits:2011zh}. In the small-error limit, this gives $\sigma_p^{\rm VLBI}\simeq15(L_p/1\,{\rm kpc})^2\,{\rm pc}$.

For the 19 binary pulsars with the required orbital-timing measurements available \citep{Egleston:2026mcg}, we estimate how their kinematic-distance uncertainties could improve after an additional 20 yr of timing. The Shklovskii contribution to the observed orbital-period derivative is obtained from $\dot P_b^{\rm Shk}=\dot P_b^{\rm obs}-\dot P_b^{\rm GW}-\dot P_b^{\rm Gal}$, where $\dot P_b^{\rm obs}$ is the observed orbital-period derivative, $\dot P_b^{\rm GW}$ is the intrinsic orbital decay due to gravitational-wave emission, and $\dot P_b^{\rm Gal}$ is the contribution from Galactic acceleration. The Shklovskii term is related to the pulsar distance through $\dot P_b^{\rm Shk}=\mu^2L_pP_b/c$, where $\mu$ is the proper motion \citep{Shklovskii70,Reardon:2024rdv}. For a pulsar with current timing baseline $T_p$, we extrapolate the published measurement uncertainties to $T_p+20\,{\rm yr}$, assuming $\sigma_{\dot P_b}\propto T^{-5/2}$ \citep{Reardon:2024rdv} and $\sigma_\mu\propto T^{-3/2}$, while keeping the uncertainties of the remaining binary parameters fixed.

To account for Galactic-potential model dependence, we adopt C20 as the fiducial model and use the larger difference from the M17 and B15 acceleration predictions at each distance as a conservative estimate of the Galactic-acceleration uncertainty \citep{Bovy:2014vfa,McMillan:2016jtx,Cautun:2019eaf}. We assume no improvement in Galactic-potential modeling over the 20-yr extrapolation and propagate this contribution together with the timing-parameter uncertainties as a Gaussian nuisance term. The resulting distance distribution defines $\sigma_p^{\rm kin}$.

Finally, we adopt
\begin{equation}
\sigma_p=
\min\left(
\sigma_p^{\rm TP},
\sigma_p^{\rm VLBI},
\sigma_p^{\rm kin}
\right),
\label{eq:distance_prior}
\end{equation}
with the kinematic term included only for these 19 binary pulsars. We do not combine the three estimates, adopting this simple approach to construct the pulsar-distance priors.
\par

\section{Analytic Relation between Pulsar Distance and the Pulsar Phase Parameter}\label{app:phip-dp}

Here we provide the explicit expression for the analytic mapping between the pulsar distance $L_p$ and the pulsar phase parameter $\Phi_p$ used in the main text. The unwrapped pulsar phase parameter can be written as a function of $L_p$ using the binary phase evolution and the Earth--pulsar light-travel time~\citep{Kato:2025set}. Specifically, for a given pulsar distance $L_p$, the unwrapped phase $\Phi_p^{\rm uw}$ is given by
\begin{align}
\Phi_p^{\rm uw}(L_p,f_0,\mathcal{M}) &= \frac{1}{32\mathcal{M}^{5/3}(\pi f_0)^{5/3}}\left[1-\left(1+\frac{256}{5}(1+\hat{\Omega}\boldsymbol{\cdot}\hat{p})\mathcal{M}^{5/3}(\pi f_0)^{8/3}L_p\right)^{5/8}\right].
\label{eq:S1}
\end{align}
Here $\Phi_p^{\rm uw}$ denotes the unwrapped phase offset of the pulsar term relative to the Earth term. Since the waveform depends on the pulsar phase parameter through $2\Phi_p$, $\Phi_p$ is physically defined modulo $\pi$. The inverse relation below is therefore obtained by setting $\Phi_p^{\rm uw}=\Phi_p-k\pi$. Analytically inverting Equation~\eqref{eq:S1} with respect to $L_p$ gives
\begin{align}
L_p(\Phi_p-k\pi,f_0,\mathcal{M}) &= \frac{5}{256}\frac{\left[1-32\mathcal{M}^{5/3}(\pi f_0)^{5/3}(\Phi_p-k\pi)\right]^{8/5}-1}{(1+\hat{\Omega}\boldsymbol{\cdot}\hat{p})\mathcal{M}^{5/3}(\pi f_0)^{8/3}}.
\label{eq:S2}
\end{align}
Here $k=0,1,2,\ldots$ indexes the multiple distance solutions arising from the $\pi$-periodicity of the pulsar phase parameter, $\hat{\Omega}$ is the unit vector pointing from the GW source to the Solar System barycenter, and $\hat{p}$ is the unit vector pointing from the Solar System barycenter to the pulsar.

For a given pulsar, evaluating Equation~\eqref{eq:S2} for each posterior sample yields a discrete set of distance solutions corresponding to different $k$ branches. These branches can give rise to multiple peaks in the posterior for $L_p$, whose widths depend on the uncertainties in the parameters entering the mapping.

\section{Dependence of Pulsar-Distance Correlations on Source Parameters}\label{app:correlations}

The posterior distributions of pulsar distances exhibit degenerate bands in the $(L_p,L_q)$ plane. These bands arise because uncertainties in the source parameters shift the inferred distances of both pulsars in a correlated manner, as illustrated in Fig.~\ref{fig_2}. In this appendix, we analyze how the slopes of these bands depend on the GW source parameters.

The two dominant sources of distance correlation are the uncertainties in $f_0$ and $\mathcal{M}$. Differentiating Equation~\eqref{eq:S2} with respect to each parameter at fixed phase $\Phi_i-k_i^*\pi$, where $k_p^*$ and $k_q^*$ are the integers that reproduce the injected values of $L_p$ and $L_q$, gives
\begin{equation}
S^{(f_0)}\equiv\frac{\partial L_q/\partial f_0}{\partial L_p/\partial f_0}=\rho\frac{1-(1+X_q)^{3/8}}{1-(1+X_p)^{3/8}},\qquad S^{(\mathcal{M})}\equiv\frac{\partial L_q/\partial\mathcal{M}}{\partial L_p/\partial\mathcal{M}}=\rho\frac{3X_q+8-8(1+X_q)^{3/8}}{3X_p+8-8(1+X_p)^{3/8}},
\label{eq:S3}
\end{equation}
where $\rho\equiv(1+\hat{\Omega}\boldsymbol{\cdot}\hat{p})/(1+\hat{\Omega}\boldsymbol{\cdot}\hat{q})$ and $X_i\equiv(256/5)(1+\hat{\Omega}\boldsymbol{\cdot}\hat{e}_i)\mathcal{M}^{5/3}(\pi f_0)^{8/3}L_i$. Here $\hat{e}_i$ is the unit vector toward pulsar $i$, such that $\hat{e}_p=\hat{p}$ and $\hat{e}_q=\hat{q}$.

To quantify the relative effects of the two parameters, we compare the first-order distance shifts induced by their uncertainties, $|\partial L_i/\partial f_0|\sigma_{f_0}$ and $|\partial L_i/\partial\mathcal{M}|\sigma_{\mathcal{M}}$, where $\sigma_{f_0}$ and $\sigma_{\mathcal{M}}$ are the corresponding posterior uncertainties. For the simulated CGW sources, the contribution from $f_0$ exceeds that from $\mathcal{M}$ in approximately 87.5\% of the evaluated pulsar--source combinations.

This result may partly arise from the slow evolution of low-frequency CGW sources, which reduces the effect of chirp-mass uncertainty on the inferred pulsar distances, and from the externally determined luminosity distances, which help constrain the chirp mass through the GW amplitude.

When the distance shifts are dominated by the uncertainty in $f_0$ and $X_p,X_q\ll1$, the slope becomes $S^{(f_0)}\approx L_q/L_p$, which is nearly independent of $f_0$ and $\rho$. Sources at different frequencies and sky positions therefore produce degenerate bands with nearly identical slopes.

\section{Pulsar-Distance Constraints from Ten CGW Sources}
\label{app:ten_sources}
\begin{figure*}[htbp]
\centering
\includegraphics[width=\textwidth]{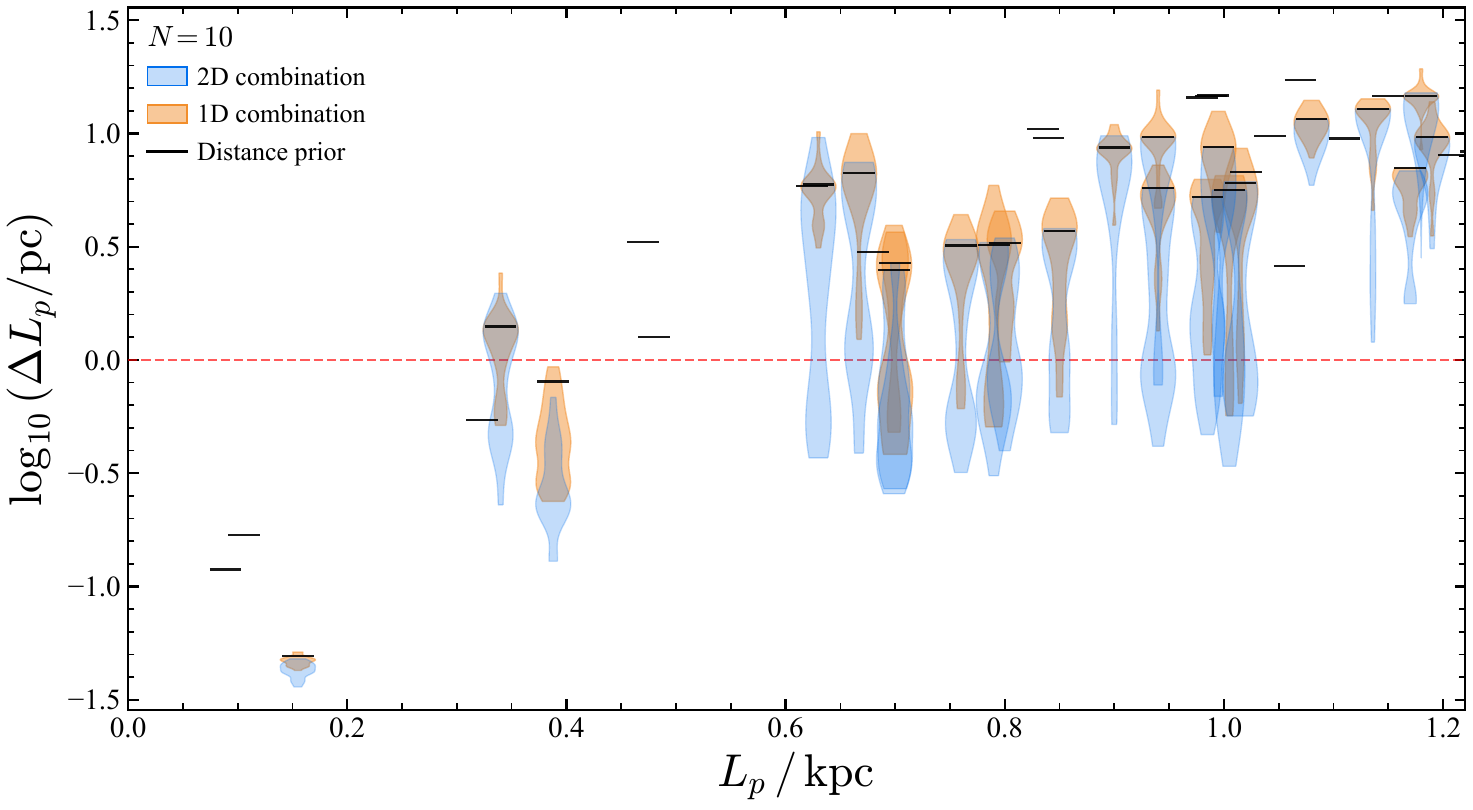}
\caption{Same as Figure~\ref{fig:dp_err}, but for $N=10$ CGW sources.}
\label{fig:dp_err_n10}
\end{figure*}

Figure~\ref{fig:dp_err_n10} presents the distributions of the pulsar-distance uncertainty $\Delta L_p$ across 20 realizations, each combining information from $N=10$ CGW sources. Compared with the $N=5$ results in Figure~\ref{fig:dp_err}, the 2D combination provides further improvements in distance precision for several pulsars.

\end{document}